\documentclass[tinyml]{acmart}
\usepackage{makecell}
\usepackage{caption}
\usepackage{subcaption}

\AtBeginDocument{%
  \providecommand\BibTeX{{%
    \normalfont B\kern-0.5em{\scshape i\kern-0.25em b}\kern-0.8em\TeX}}}

\setcopyright{rightsretained}
\copyrightyear{2025}
\acmYear{2025}

\begin{document}

\title[\resizebox{4.5in}{!}{A 28 nm AI microcontroller with tightly coupled zero-standby power weight memory featuring standard logic compatible 4 Mb 4-bits/cell embedded flash technology}]{A 28 nm AI microcontroller with tightly coupled zero-standby power weight memory featuring standard logic compatible 4 Mb 4-bits/cell embedded flash technology}

\author{Daewung Kim}
\affiliation{%
  \institution{ANAFLASH Inc.}
  \streetaddress{169 Yeoksam-ro, Gangnam-gu}
  \city{Seoul}
  \country{Republic of Korea}
  \postcode{06247}
}
\email{david@anaflash.com}

\author{Seong Hwan Jeon}
\affiliation{%
  \institution{ANAFLASH Inc.}
  \streetaddress{169 Yeoksam-ro, Gangnam-gu}
  \city{Seoul}
  \country{Republic of Korea}
  \postcode{06247}
}
\email{john@anaflash.com}

\author{Young Hee Jeon}
\affiliation{%
  \institution{ANAFLASH Inc.}
  \streetaddress{169 Yeoksam-ro, Gangnam-gu}
  \city{Seoul}
  \country{Republic of Korea}
  \postcode{06247}
}
\email{nina@anaflash.com}

\author{Kyung-Bae Kwon}
\affiliation{%
  \institution{ANAFLASH Inc.}
  \streetaddress{169 Yeoksam-ro, Gangnam-gu}
  \city{Seoul}
  \country{Republic of Korea}
  \postcode{06247}
}
\email{luke@anaflash.com}

\author{Jigon Kim}
\affiliation{%
  \institution{ANAFLASH Inc.}
  \streetaddress{169 Yeoksam-ro, Gangnam-gu}
  \city{Seoul}
  \country{Republic of Korea}
  \postcode{06247}
}
\email{jerry@anaflash.com}

\author{Yeounghun Choi}
\affiliation{%
  \institution{ANAFLASH Inc.}
  \streetaddress{169 Yeoksam-ro, Gangnam-gu}
  \city{Seoul}
  \country{Republic of Korea}
  \postcode{06247}
}
\email{eun@anaflash.com}

\author{Hyunseung Cha}
\affiliation{%
  \institution{ANAFLASH Inc.}
  \streetaddress{Bundangnaegok-ro, Bundang-gu}
  \city{Seongnam-si}
  \country{Republic of Korea}
  \postcode{13529}
}
\email{tony@anaflash.com}

\author{Kitae Kwon}
\affiliation{%
  \institution{ANAFLASH Inc.}
  \streetaddress{440 N WOLFE RD}
  \city{Sunnyvale}
  \state{CA}
  \country{USA}
  \postcode{94085-3869}
}
\email{kkwon@anaflash.com}

\author{Daesik Park}
\affiliation{%
  \institution{ANAFLASH Inc.}
  \streetaddress{440 N WOLFE RD}
  \city{Sunnyvale}
  \state{CA}
  \country{USA}
  \postcode{94085-3869}
}
\email{daniel@anaflash.com}

\author{Jongseuk Lee}
\affiliation{%
  \institution{ANAFLASH Inc.}
  \streetaddress{440 N WOLFE RD}
  \city{Sunnyvale}
  \state{CA}
  \country{USA}
  \postcode{94085-3869}
}
\email{jimmy@anaflash.com}

\author{Sihwan Kim}
\affiliation{%
  \institution{ANAFLASH Inc.}
  \streetaddress{440 N WOLFE RD}
  \city{Sunnyvale}
  \state{CA}
  \country{USA}
  \postcode{94085-3869}
}
\email{skim@anaflash.com}

\author{Seung-Hwan Song}
\affiliation{%
  \institution{ANAFLASH Inc.}
  \streetaddress{Bundangnaegok-ro, Bundang-gu}
  \city{Seongnam-si}
  \country{Republic of Korea}
  \postcode{13529}
}
\affiliation{%
  \streetaddress{440 N WOLFE RD}
  \city{Sunnyvale}
  \state{CA}
  \country{USA}
  \postcode{94085-3869}
}
\email{peter@anaflash.com}

\renewcommand{\shortauthors}{Kim, et al.}

\begin{abstract}
This study introduces a novel AI microcontroller optimized for cost-effective, battery-powered edge AI applications. Unlike traditional single bit/cell memory configurations, the proposed microcontroller integrates zero-standby power weight memory featuring standard logic compatible 4-bits/cell embedded flash technology tightly coupled to a Near-Memory Computing Unit. This architecture enables efficient and low-power AI acceleration. Advanced state mapping and an overstress-free word line (WL) driver circuit extend verify levels, ensuring robust 16 state cell margin. A ping-pong buffer reduces internal data movement while supporting simultaneous multi-bit processing. The fabricated microcontroller demonstrated high reliability, maintaining accuracy after 160 hours of unpowered baking at 125℃.
\end{abstract}

\keywords{Non-volatile memory, Near-Memory Compute, Microcontroller}

\maketitle

\section{Introduction}
Microcontrollers designed for battery-powered smart edge devices are often required to run inferencing tasks at a place where sensor data are generated for real-time response. They use a locally stored AI model which is trained in the cloud.
Power-gating technique is often deployed to reduce idle mode power consumption in the low power applications. The AI model can be stored and updated in an embedded Non-Volatile Memory (eNVM) during the device's lifetime without consuming standby power during the idle mode.
Typically, multiple-time programmable eNVM technology requires additional fabrication steps beyond a standard logic process and are configured to store only single bit information per unit memory cell, which limits the efficiency of AI computation \cite{Deaville22}.
In this work, we introduce an AI microcontroller with zero-standby power weight memory featuring standard logic compatible 4-bits/cell Embedded FLASH (EFLASH) technology, tightly coupled with a Near-Memory Computing Unit (NMCU) for cost-effective and low power edge AI computing applications.

\section{The Proposed Architecture}
\subsection{Overall Structure}
Fig. 1 shows a block diagram of the proposed AI microcontroller, which consists of i) 32-bit RISC-V CPU core, ii) SRAM for instruction and data memory, iii) DMA controller, iv) peripheral subsystems including GPIO, SPI, and UART, v) 128 Kb EFLASH for initial setting parameters and code storage, vi) 4 Mb 4-bits/cell EFLASH tightly coupled with the NMCU, v) on-chip standard logic compatible High Voltage (HV) and reference voltage generator circuits. The EFLASH macro is based on a 5T cell based single-poly EFLASH cell array\cite{Song13} and is integrated with other peripheral circuits such as in the WL driver, Sense Amplifier (SA) circuits.
\begin{figure}[ht]
  \centering
  \includegraphics[width=\linewidth]{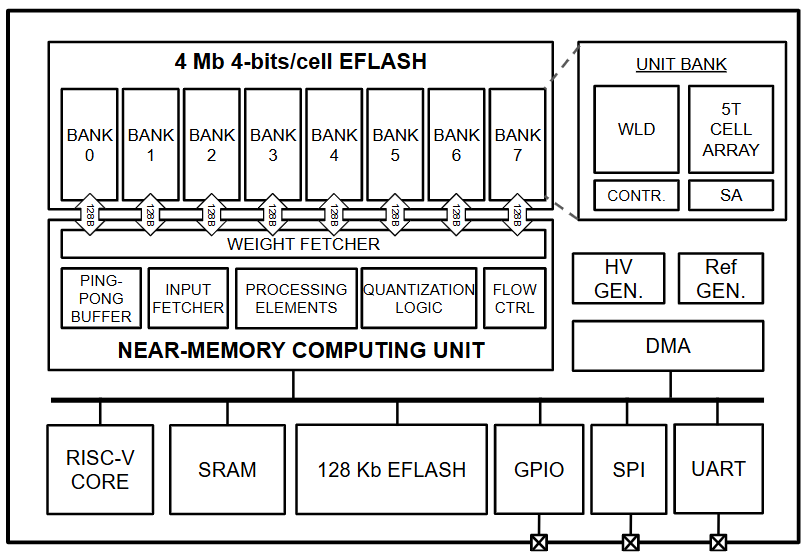}
  \caption{AI microcontroller featuring 4-bits/cell EFLASH technology tightly coupled to a near-memory computing unit}
  \Description{AI microcontroller featuring 4-bits/cell EFLASH technology tightly coupled to a near-memory computing unit}
\end{figure}

\subsection{Near-Memory Computing Unit(NMCU)}
\begin{figure*}[!ht]
  \centering
  \includegraphics[width=0.8\textwidth]{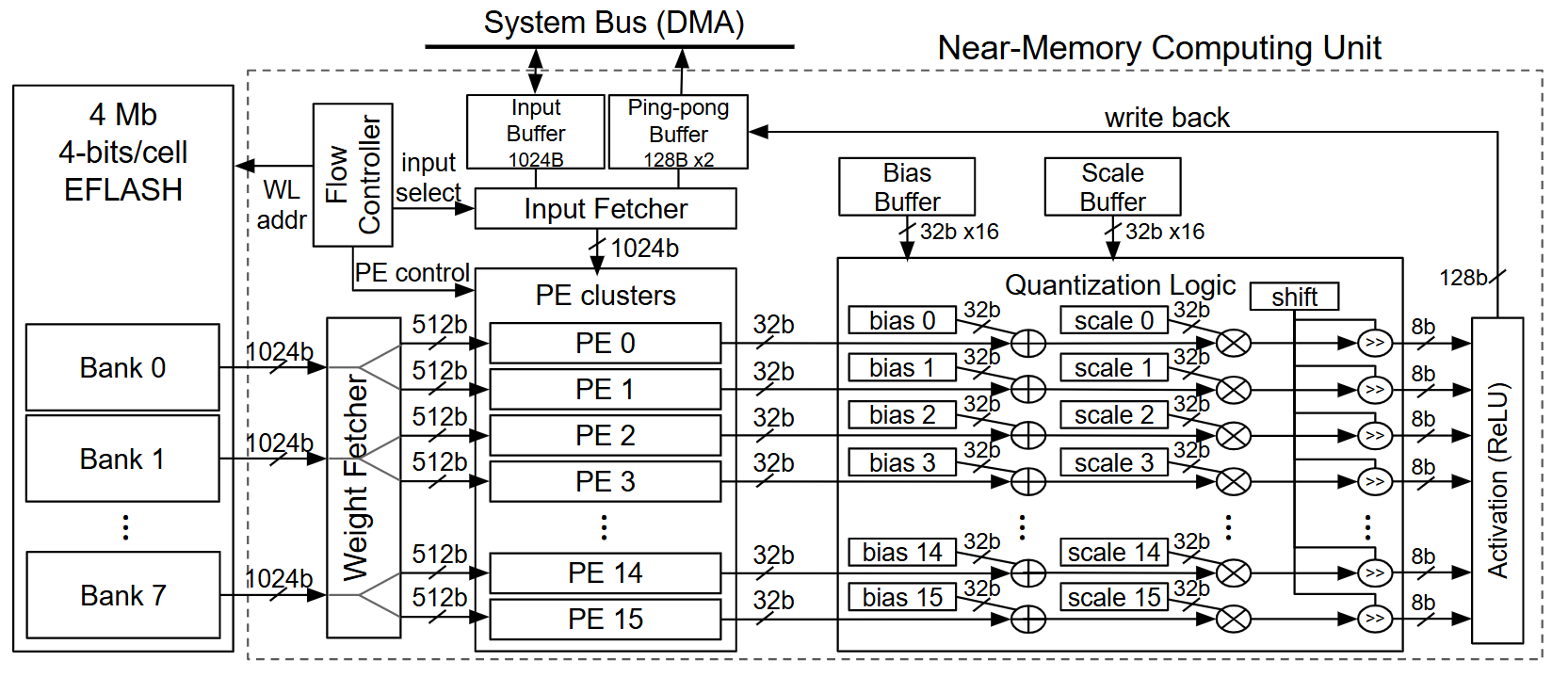}
  \caption{Near-Memory Computing Unit for efficient AI acceleration}
  \Description{Near-Memory Computing Unit for efficient AI acceleration}
\end{figure*}
Fig. 2 shows a Near-Memory Computing Unit with 4-bits/cell EFLASH based weight memory. The flash memory is tightly coupled to the computation unit with a large bandwidth for efficient AI acceleration. Each 4-bits/cell EFLASH bank can load 256 4-bit weights in a single read operation. To maximize throughput, two processing elements (PEs) are allocated per 4-bits/cell EFLASH macro. Therefore, one PE can process MAC operations of up to 128 elements per EFLASH read. Larger matrix-vector multiplication (MVM) operations are possible by performing multiple EFLASH reads in succession. The NMCU's flow control logic automatically adjusts the address of the weight parameters as required for the MVM operation with a single RISC-V instruction, which reduces communication overhead between host CPU and NMCU. NMCU includes a ping-pong buffer that can use the calculation results of the previous layer as an input for the next layer calculation. The input fetcher logic supplies the PE with an input vector of 128 8-bit elements by selecting either the input buffer or the ping-pong buffer. After the MVM operation is completed, the operation result is quantized to 8 bits and written-back to the ping-pong buffer. Notably, no additional data movement is required beyond the first input vector for TinyML models like FC-Autoencoder \cite{Banbury21}. The NMCU also employs element-wise int8 quantization schemes from TFLite-micro \cite{Jacob18}.

\subsection{High Voltage Generator}
Fig. 3 shows the schematic diagram of the designed HV generator circuit to pump the I/O supply voltage (i.e. VDDH = 2.5V) to program and erase voltage level (i.e. VPP4 = $\sim$10 V) during the program and erase operations. The HV generator is designed using standard I/O logic devices without any additional HV process steps and is composed of six-stage voltage doubler to operate the individual I/O devices within the nominal operating voltage level while providing sufficiently high regulated VPP4 level for given program and erase time. Here, we deploy the adaptive body biasing scheme for NMOS as well as PMOS transistors to avoid forward bias current in the voltage doubler circuit. When the VPP1 level is boosted higher than a reference level of SREF, the cascaded PMOS switches connect the boosted nodes VPP1-4 to the program/erase voltage supply nodes (i.e. VPS1-4) without introducing stress voltage of the PMOS switches during the program/erase operation of the logic compatible EFLASH macro. On the other hand, when the VPP1 level is discharged lower than a reference level of SREF by disabling the clock generator to save power consumption from the HV generator circuit, the cascaded PMOS switches connect the VDDH level to the program/erase voltage supply nodes VPS1-4.
\begin{figure}[ht]
  \centering
  \includegraphics[width=\linewidth]{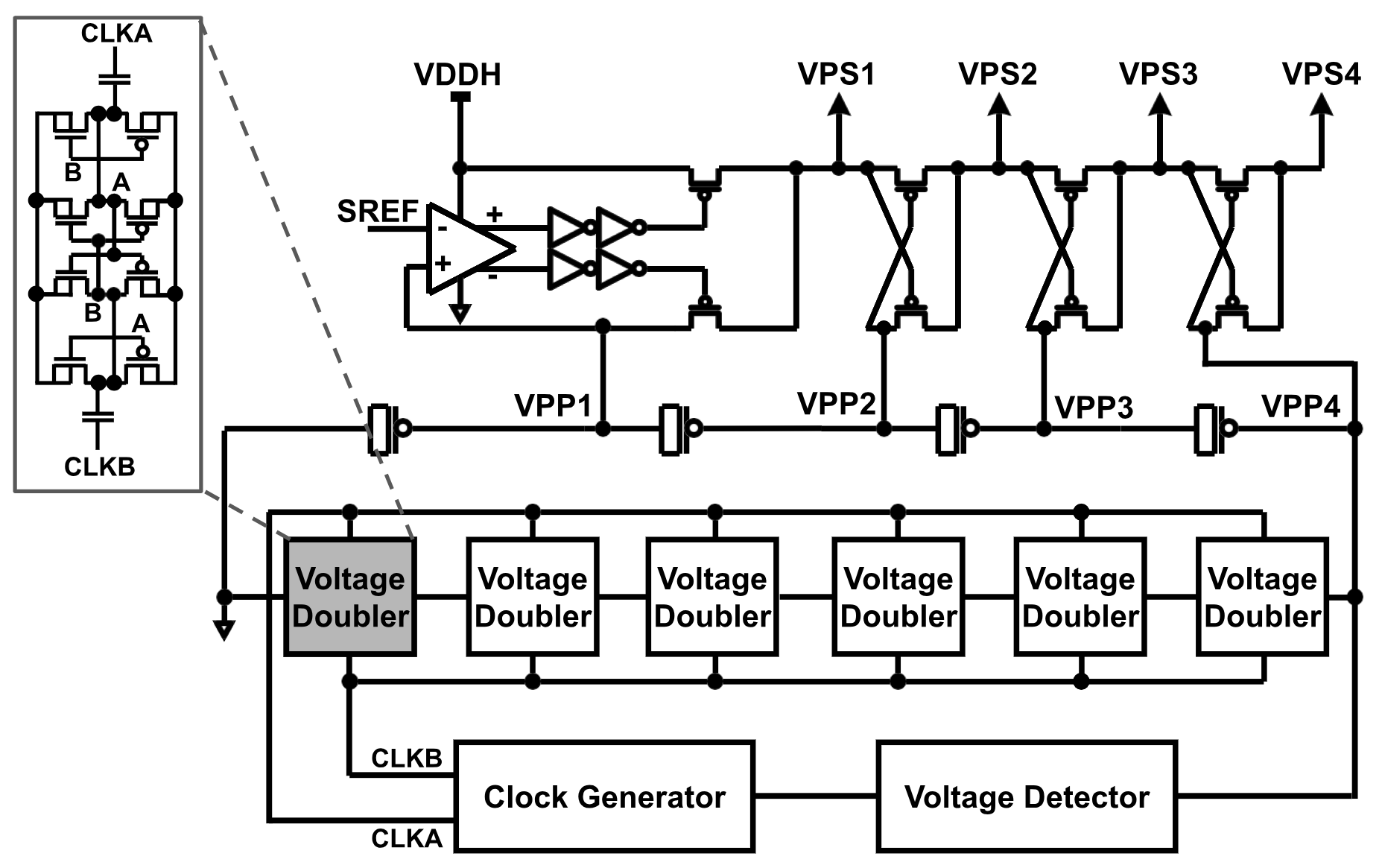}
  \caption{Standard logic compatible high voltage generator for embedded flash program/erase operations}
  \Description{Standard logic compatible high voltage generator for embedded flash program/erase operations}
\end{figure}

\subsection{Overstress-free WL driver}
The conventional WL driver circuit in \cite{Song13} supplies a read reference level (i.e. VRD) through the source of the NMOS device string to the selected WL. Due to the threshold voltage drop of the NMOS exacerbated by the elevated source voltage of the NMOS, the available VRD for the EFLASH read operation was much lower than VPPH level. In this work, we propose an overstress-free WL driver circuit with a VRD PMOS charging path from the PMOS charging circuit as shown in Fig. 4. This driver is used to extend the VRD level up to VDDH (nominal operating voltage of the individual device) for a wider range program-verify read operation, which is critical for 4-bits/cell program verify operations. For program operation, SWR1 and SWR2 signals are toggled. Then, WL can be driven to the program voltage (i.e. VPGM=10 V) through the VPGM charging PMOS path shown in Fig 4a. Since the stacked devices in the VPGM discharging path split the voltage stresses, the driver circuit operates without introducing voltage overstress of the individual device. For program-verify operation, the read selection signal (SRD) is switched from low to high. Then, as illustrated in Fig 4b, the WL starts charging from GND to the VRD level through the VRD NMOS path for the case when the VRD is low enough and through the VRD PMOS path for the case when the VRD is high enough. When the SRD is switched from high to low, the WL is connected to the ground level through the NMOS discharging path. Thus, with the proposed circuit, the program-verify read voltage of VRD can be extended to VDDH without a VTH drop. For read operation, the high voltage generator circuit is turned off. Then, VPS1-4 nodes are switched to VDDH, whereas VPP1-4 nodes are switched to GND from the circuits shown in Fig. 3. Then, as illustrated in Fig 4c, the WL begins charging from GND to the VRD level through the VRD NMOS and/or PMOS path depending on the VRD level. Consequently, the proposed circuit extends the read voltage of VRD to VDDH without a VTH drop, enabling reliable 4-bit/cell read operation.

\begin{figure}
  \centering
  \begin{subfigure}[b]{0.5\textwidth}
    \centering
    \includegraphics[width=\textwidth]{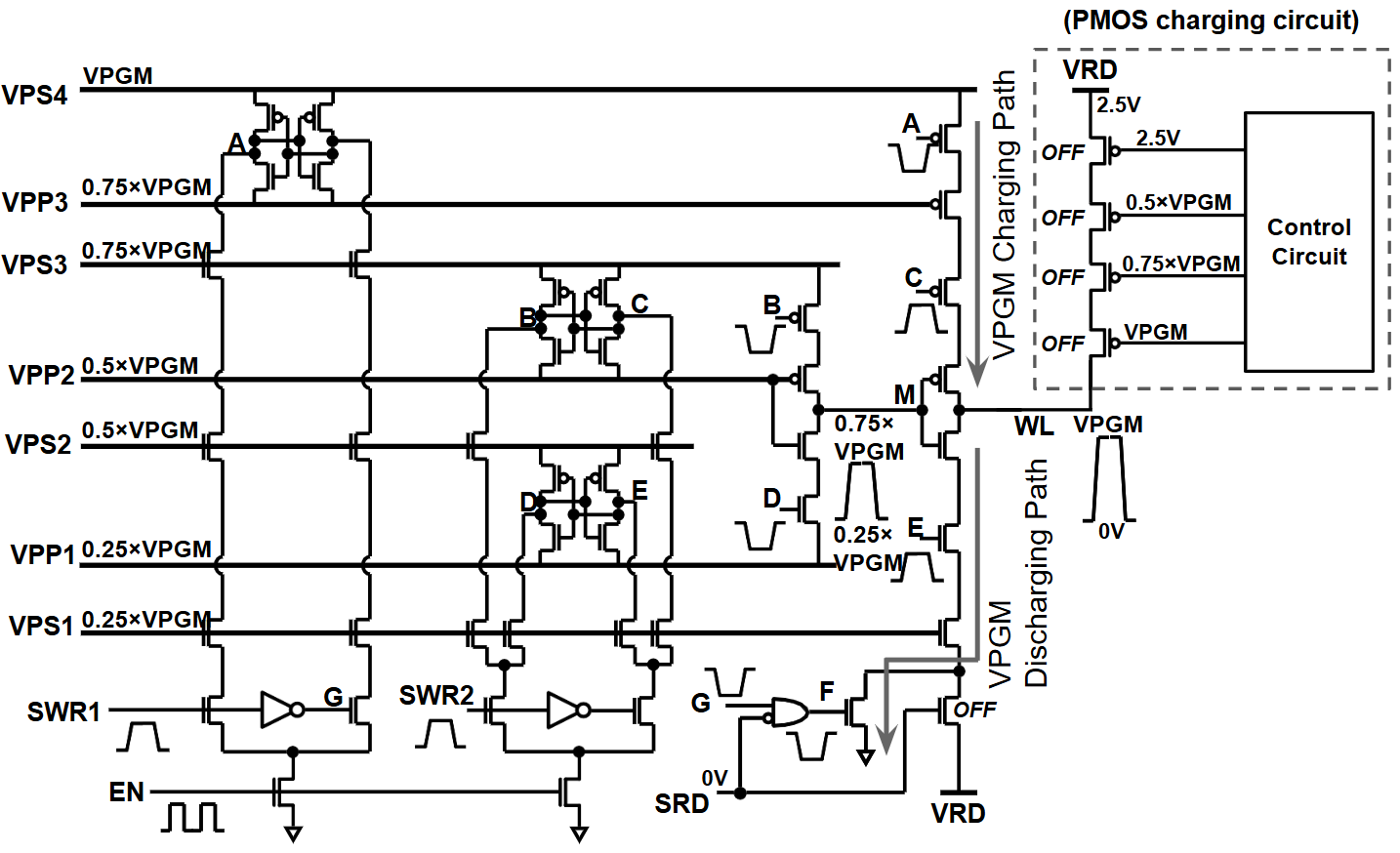}
    \caption{Program operation}
  \end{subfigure}
  \begin{subfigure}[b]{0.5\textwidth}
    \centering
    \includegraphics[width=\textwidth]{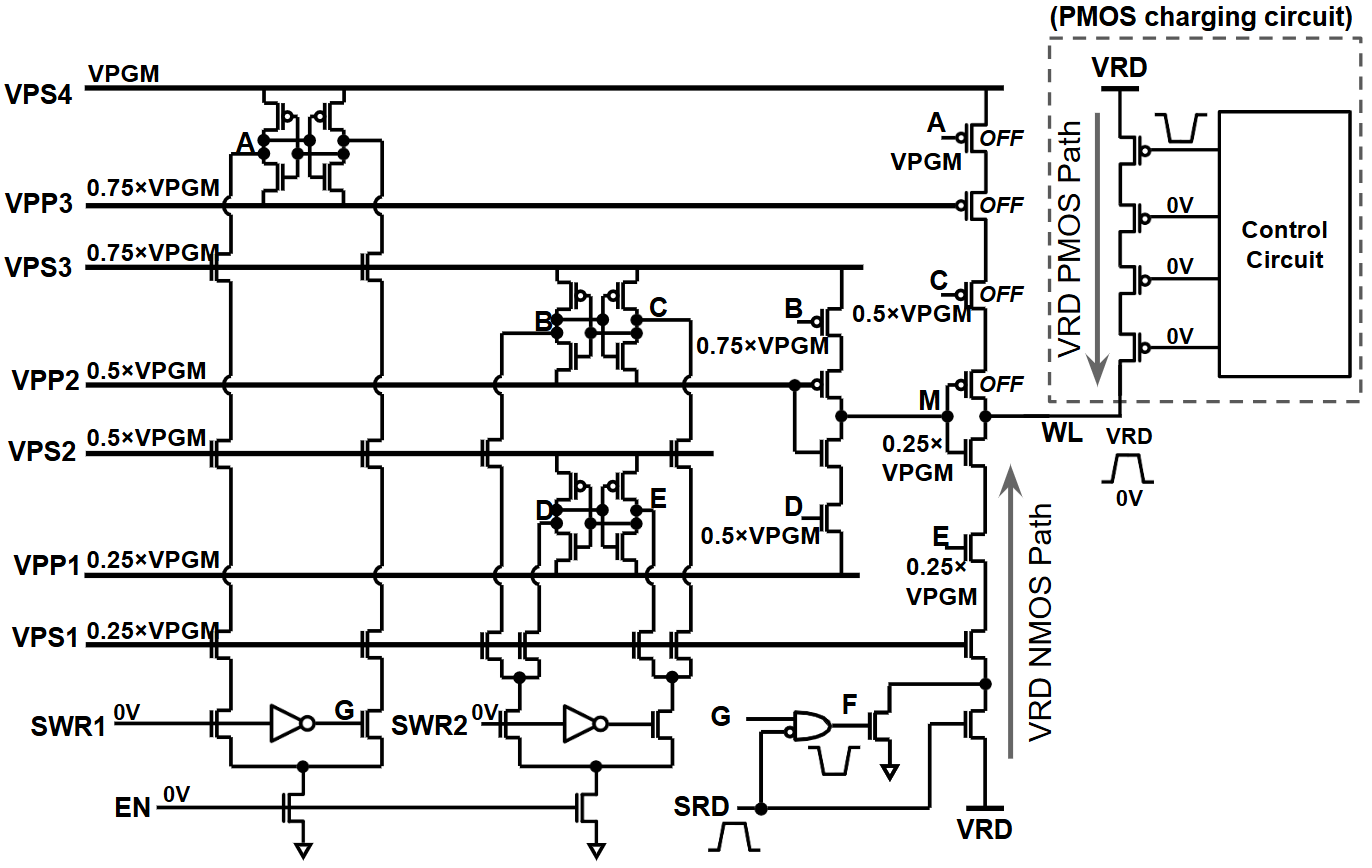}
    \caption{Program-verify operation}
  \end{subfigure}
  \begin{subfigure}[b]{0.5\textwidth}
    \centering
    \includegraphics[width=\textwidth]{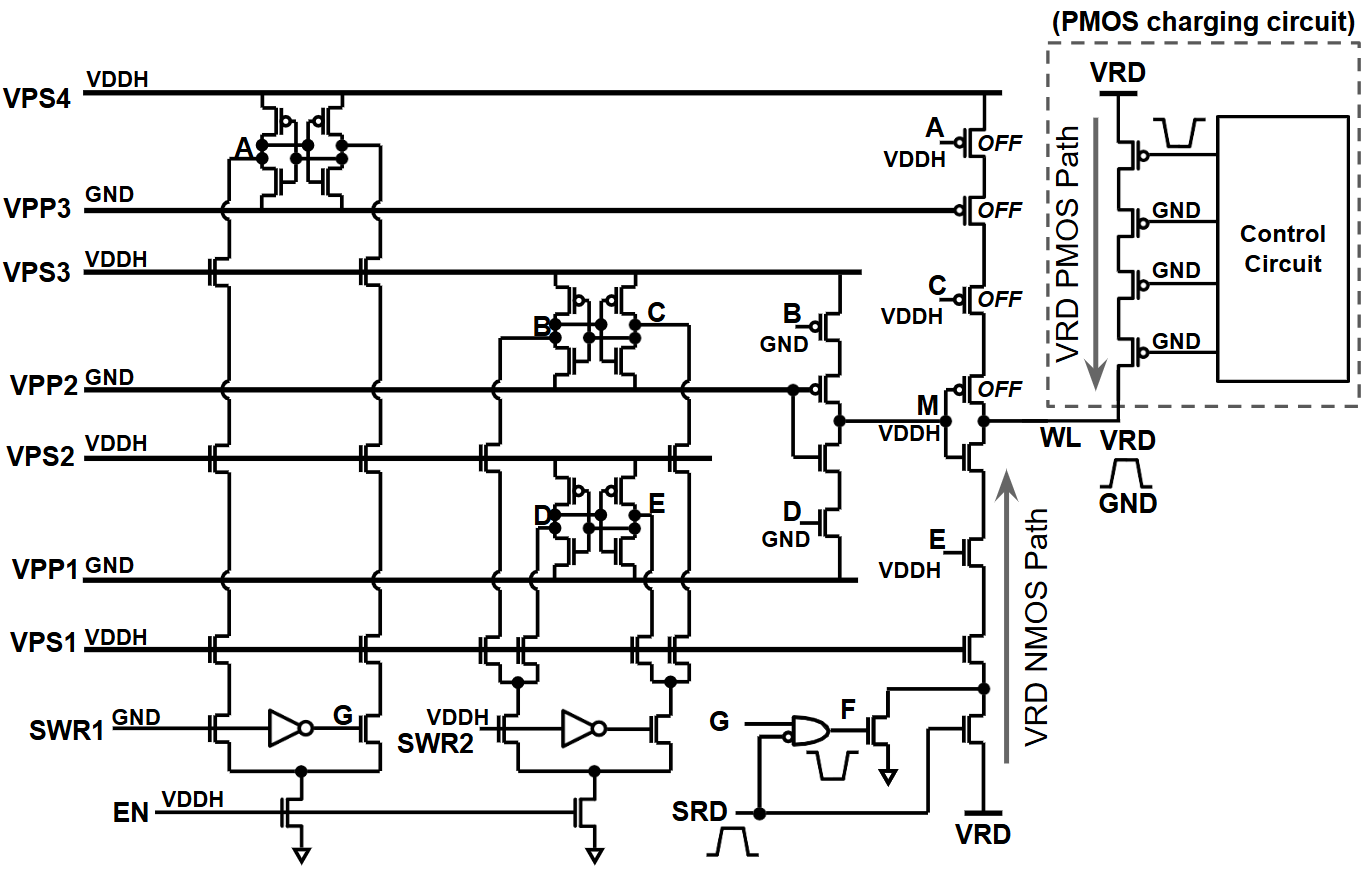}
    \caption{Read operation}
  \end{subfigure}
    \caption{Overstress-free WL driver circuit of 4-bits/cell EFLASH with PMOS charging path: (a) for program operation, (b) for a program-verify read operation, and (c) for read operation.}
    \Description{Overstress-free WL driver circuit of 4-bits/cell EFLASH with PMOS charging path: (a) for program operation, (b) for a program-verify read operation, and (c) for read operation.}
\end{figure}

\section{Experiment Results}
The proposed standard logic compatible non-volatile AI microcontroller featuring 4-bits/cell EFLASH tightly coupled to the NMCU has been fabricated using a 1 V core supply 28 nm low power standard logic technology. Since the 4-bits/cell EFLASH cells have a higher probability of transitioning to the adjacent states compared to the long distance states during a cell lifetime, we mapped the 4-bits/cell EFLASH memory states to the 4-bit quantized weight value such that the adjacent states can differ by one decimal value as shown in Fig. 5 (a). This resulted in a non-uniform distribution of the programmed 4-bits/cell EFLASH memory states, since the distribution of the trained weights is the most common near zero in general \cite{Zhong22}. Considering such a non-uniform distribution, we carefully determined 15 verify read reference levels for 15 programmed states. By sequentially verifying each programmed state as shown in Fig.5 (b), 16 distinct states can be programmed with a margin between states. The designed logic compatible HV generator circuits were measured to boost the program voltage level (i.e. VPP4 level) of approximately 10V as shown in Fig. 5 (c). The designed WL driver circuits were measured to supply verify-reference levels from 0V to 2.5V (=VDDH), which is used to verify 15 programmed states with a full range of 2.5V.

\begin{figure}
  \centering
  \begin{subfigure}[b]{0.4\textwidth}
    \centering
    \includegraphics[width=\textwidth]{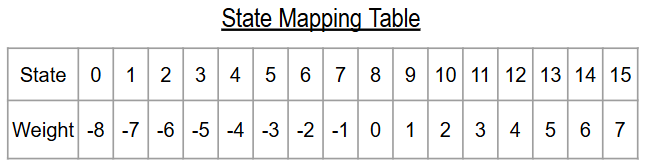}
    \caption{}
  \end{subfigure}
  \begin{subfigure}[b]{0.4\textwidth}
    \centering
    \includegraphics[width=\textwidth]{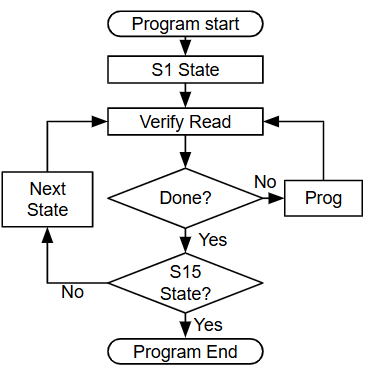}
    \caption{}
  \end{subfigure}
  \begin{subfigure}[b]{0.4\textwidth}
    \centering
    \includegraphics[width=\textwidth]{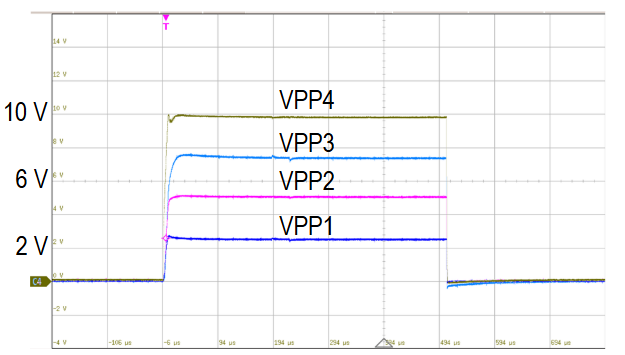}
    \caption{}
  \end{subfigure}
  \begin{subfigure}[b]{0.4\textwidth}
    \centering
    \includegraphics[width=\textwidth]{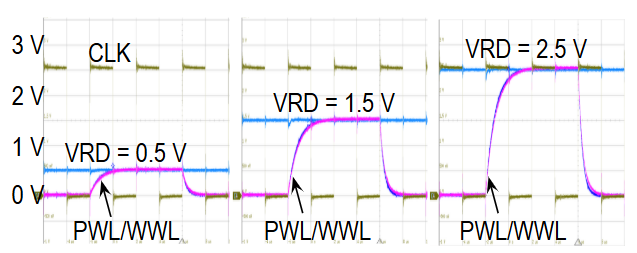}
    \caption{}
  \end{subfigure}
    \caption{(a) 4-bits/cell EFLASH state mapping table, (b) 16 states program-verify sequences, (c) measured VPP1-4 levels from the logic compatible charge pump, and (d) WL driver output signals (PWL/WWL) for verify operations of 4-bits/cell EFLASH cells}
    \Description{(a) 4-bits per cell EFLASH state mapping table, (b) 16 states program-verify sequences, (c) measured VPP1-4 levels from the logic compatible charge pump, and (d) WL driver output signals (PWL and WWL) for verify operations of 4-bits per cell EFLASH cells}
\end{figure}

To demonstrate actual neural networks in our chip, we evaluated an MLP model trained with MNIST dataset \cite{LeCun98} and standard benchmark FC-Autoencoder from MLPerf-Tiny \cite{Banbury21} before and after baking the fabricated microcontroller chip at 125℃ for 340 and 160 hours, respectively. To fit the precision of the weights to 4 bits/cell EFLASH, we performed 4 bit integer quantization aware training with MNIST dataset and ToyADMOS dataset. Fig. 6 shows the measured weight distribution of 4-bits/cell EFLASH cells and Table~\ref{tab:infer} shows AI inference test results. Although some overlap was observed between adjacent cell states after baking, AI inference accuracy remained robust to have 95.58\% for MNIST and 0.878 AUC for FC-Autoencoder, respectively. As a result, the inference accuracy degradation was limited to 0.04\% compared to the software baseline for the MNIST dataset or not observed for FC-Autoencoder dataset for which the 9th layer of the model was implemented on-chip while other layers were processed off-chip as described in Fig. 7.

\begin{figure}[!ht]
  \centering
  \begin{subfigure}[b]{0.42\textwidth}
    \centering
    \includegraphics[width=\textwidth]{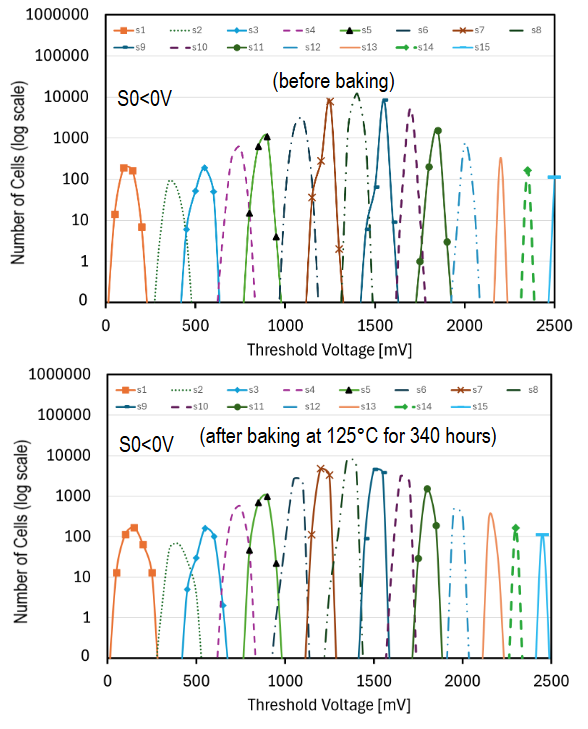}
    \caption{Weight distribution for MNIST(34K cells)}
  \end{subfigure}
  \par\bigskip
  \begin{subfigure}[b]{0.42\textwidth}
    \centering
    \includegraphics[width=\textwidth]{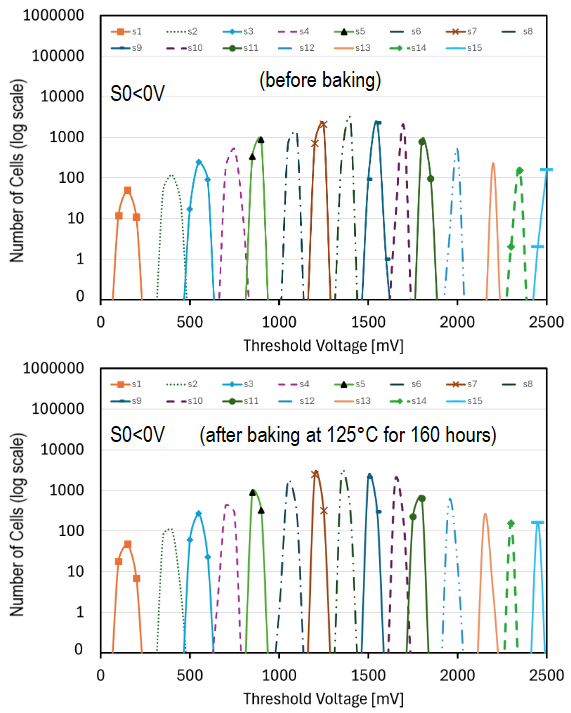}
    \caption{Weight distribution for Autoencoder(16K cells)}
  \end{subfigure}
    \caption{Measured weight distribution of 4-bits/cell EFLASH cells}
    \Description{Measured weight distribution of 4-bits per cell EFLASH cells}
\end{figure}

\begin{table}
  \caption{Measured results of AI inference tasks}
  \label{tab:infer}
  \begin{tabular}{ccl}
    \toprule
    Inference Accuracy & MNIST & AutoEncoder\\
    \midrule
    Before Bake &  95.67\% & 0.878 AUC\\
    After Bake & 95.58\% & 0.878 AUC\\
    SW. Baseline & 95.62\% & 0.878 AUC\\
  \bottomrule
\end{tabular}
\end{table}

\begin{figure}[ht]
  \centering
  \includegraphics[width=\linewidth]{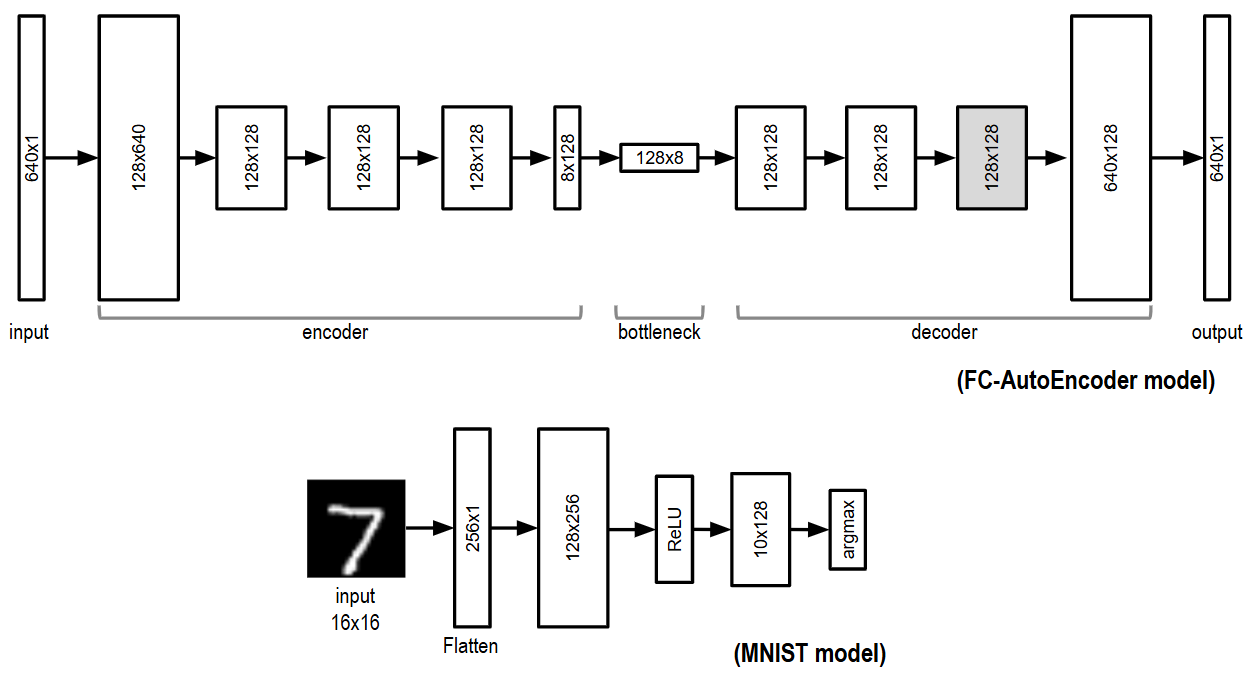}
  \caption{AI inference model}
  \Description{AI inference model}
\end{figure}

\section{Conclusion}
As summarized in the Table \ref{tab:comparison}, this work presents a unique standard logic compatible non-volatile microcontroller designed for cost-effective battery-powered edge AI device applications. A die photograph of the fabricated AI microcontroller is shown in Fig. 8. While alternative AI acceleration solutions employing tightly coupled memory have largely been restricted to a single bit/cell configurations, the proposed AI microcontroller, with its tightly coupled zero-standby-power weight memory, incorporates standard logic compatible 4-bit/cell embedded flash technology for efficient low power edge AI acceleration. Carefully designed state mapping and overstress-free WL driver circuit provide wider-range of verify levels, enabling a sufficient cell margin for 16 distinct cell states. The tightly coupled NMCU processes multi-bit information simultaneously and minimizes an internal data movement by a carefully designed ping-pong buffer. The fabricated non-volatile AI microcontroller maintained a good accuracy after being baked at 125℃ for more than 160 hours while unpowered.
\begin{figure}[ht]
  \centering
  \includegraphics[width=0.7\linewidth]{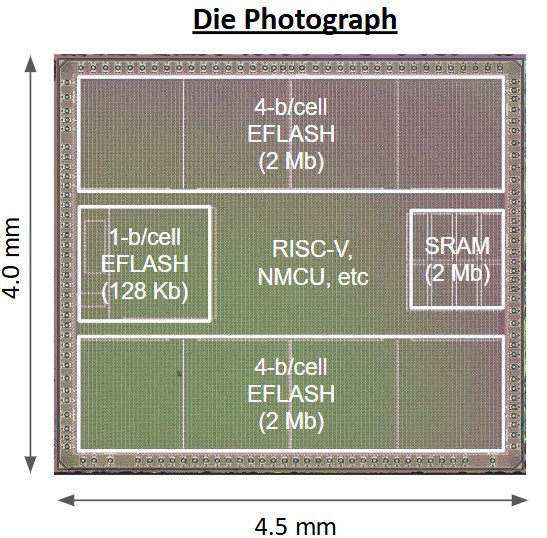}
  \caption{Die photograph of the fabricated AI microcontroller}
  \Description{Die photograph of the fabricated AI microcontroller}
\end{figure}

\begin{table}
  \caption{Comparison table}
  \label{tab:comparison}
  \begin{tabular}{ccccc}
    \toprule
     & \cite{Deaville22} & \cite{Desoli23} & \cite{Lin23} & This Work\\
    \midrule
    Process & 22 nm &  18 nm & 28 nm & 28 nm\\
    Process Overhead & Yes & No & No & No\\
    Memory Config & \makecell{1 bit/cell \\ MRAM} & \makecell{1 bit/cell \\ SRAM} & \makecell{1 bit/cell \\ SRAM} & \makecell{4 bits/cell \\ EFLASH}\\
    Non-Volatile & Yes & No & No & Yes\\
    Activation Precision & 1b & 1-4b & 8b & 8b\\
    Weight Precision & 4b & 1-4b & 8b & 4b\\
  \bottomrule
\end{tabular}
\end{table}

\begin{acks}
This work was partly supported by the DIPS 1000+ Fabless Challenge award and TIPS grants funded by the Ministry of SMEs and Startups (S3318962) and the Institute of Information \& Communications Technology Planning \& Evaluation (IITP) grants funded by the Korea government (MSIT) (RS-2023-00216370 and RS-2023-00229849).
\end{acks}

\bibliographystyle{ACM-Reference-Format}
\bibliography{acmart}


\begin{thebibliography}{8}


\ifx \showCODEN    \undefined \def \showCODEN     #1{\unskip}     \fi
\ifx \showDOI      \undefined \def \showDOI       #1{#1}\fi
\ifx \showISBNx    \undefined \def \showISBNx     #1{\unskip}     \fi
\ifx \showISBNxiii \undefined \def \showISBNxiii  #1{\unskip}     \fi
\ifx \showISSN     \undefined \def \showISSN      #1{\unskip}     \fi
\ifx \showLCCN     \undefined \def \showLCCN      #1{\unskip}     \fi
\ifx \shownote     \undefined \def \shownote      #1{#1}          \fi
\ifx \showarticletitle \undefined \def \showarticletitle #1{#1}   \fi
\ifx \showURL      \undefined \def \showURL       {\relax}        \fi
\providecommand\bibfield[2]{#2}
\providecommand\bibinfo[2]{#2}
\providecommand\natexlab[1]{#1}
\providecommand\showeprint[2][]{arXiv:#2}

\bibitem[Deaville et~al\mbox{.}(2022)]%
        {Deaville22}
\bibfield{author}{\bibinfo{person}{Peter Deaville}, \bibinfo{person}{Bonan
  Zhang}, {and} \bibinfo{person}{Naveen Verma}.}
  \bibinfo{year}{2022}\natexlab{}.
\newblock \bibinfo{booktitle}{\emph{A 22nm 128-kb MRAM Row/Column-Parallel
  In-Memory Computing Macro with Memory-Resistance Boosting and Multi-Column
  ADC Readout}}.
\newblock Symposium on VLSI Technology \& Circuits.
\newblock


\bibitem[et~al.(2018)]%
        {Jacob18}
\bibfield{author}{\bibinfo{person}{B.~Jacob et al.}}
  \bibinfo{year}{2018}\natexlab{}.
\newblock \bibinfo{booktitle}{\emph{Quantization and Training of Neural
  Networks for Efficient Integer-Arithmetic-Only Inference}}.
\newblock IEEE CVPR.
\newblock


\bibitem[et~al.(2021)]%
        {Banbury21}
\bibfield{author}{\bibinfo{person}{C.~Banbury et al.}}
  \bibinfo{year}{2021}\natexlab{}.
\newblock \bibinfo{booktitle}{\emph{MLPerf Tiny Benchmark}}.
\newblock Thirty-fifth Conference on Neural Information Processing Systems
  Datasets and Benchmarks Track (Round 1).
\newblock
\newblock
\shownote{\url{https://openreview.net/forum?id=8RxxwAut1BI}}.


\bibitem[et~al.(2023)]%
        {Desoli23}
\bibfield{author}{\bibinfo{person}{Desoli et al.}}
  \bibinfo{year}{2023}\natexlab{}.
\newblock \bibinfo{booktitle}{\emph{16.7 A 40-310TOPS/W SRAM-Based All-Digital
  Up to 4b In-Memory Computing Multi-Tiled NN Accelerator in FD-SOI 18nm for
  Deep-Learning Edge Applications}}.
\newblock ISSCC.
\newblock


\bibitem[LeCun et~al\mbox{.}(1998)]%
        {LeCun98}
\bibfield{author}{\bibinfo{person}{Y. LeCun}, \bibinfo{person}{L. Bottou},
  \bibinfo{person}{Y. Bengio}, {and} \bibinfo{person}{P. Haffner}.}
  \bibinfo{year}{1998}\natexlab{}.
\newblock \bibinfo{booktitle}{\emph{Gradient-based learning applied to document
  recognition}}.
\newblock Proceedings of the IEEE.
\newblock


\bibitem[Lin et~al\mbox{.}(2023)]%
        {Lin23}
\bibfield{author}{\bibinfo{person}{Chuan-Tung Lin},
  \bibinfo{person}{Paul~Xuanyuanliang Huang}, \bibinfo{person}{Jonghyun Oh},
  \bibinfo{person}{Dewei Wang}, {and} \bibinfo{person}{Mingoo Seok}.}
  \bibinfo{year}{2023}\natexlab{}.
\newblock \bibinfo{booktitle}{\emph{iMCU: A 102-$\mu$J, 61-ms Digital In-Memory
  Computing-based Microcontroller Unit for Edge TinyML}}.
\newblock CICC.
\newblock


\bibitem[Song et~al\mbox{.}(2013)]%
        {Song13}
\bibfield{author}{\bibinfo{person}{Seung-Hwan Song}, \bibinfo{person}{Ki~Chul
  Chun}, {and} \bibinfo{person}{Chris~H. Kim}.}
  \bibinfo{year}{2013}\natexlab{}.
\newblock \bibinfo{booktitle}{\emph{A Logic-Compatible Embedded Flash Memory
  for Zero-Standby Power System-on-Chips Featuring a Multi-Story High Voltage
  Switch and a Selective Refresh Scheme}}.
\newblock IEEE JSSC.
\newblock


\bibitem[Zhong et~al\mbox{.}(2022)]%
        {Zhong22}
\bibfield{author}{\bibinfo{person}{Weishun Zhong}, \bibinfo{person}{Ben
  Sorscher}, \bibinfo{person}{Daniel~D Lee}, {and} \bibinfo{person}{Haim
  Sompolinsky}.} \bibinfo{year}{2022}\natexlab{}.
\newblock \bibinfo{booktitle}{\emph{A theory of learning with constrained
  weight-distribution}}.
\newblock 36th Conference on Neural Information Processing Systems.
\newblock


\end{thebibliography}

\end{document}